\documentclass[12pt]{article}
\usepackage{amsfonts}

\makeatletter \@addtoreset{equation}{section} \makeatother

\renewcommand{\theequation}{\thesection.\arabic{equation}}

\begin{document}

\title{Backlund transformations for the Nizhnik-Novikov-Veselov equation.}
\author{V.E.Vekslerchik}
\date{\today}
\maketitle

\begin{center}
{\it
  Institute for Radiophysics and Electronics, Kharkov, Ukraine
  \\
  and
  \\
  Universidad de Castilla-La Mancha, Ciudad Real, Spain
}
\end{center}

\begin{abstract}
The Backlund transformations for the Nizhnik-Novikov-Veselov
equation are presented. It is shown that these transformations can
be iterated and that the resulting sequence can be described by
the Volterra equations. The relationships between the
Nizhnik-Novikov-Veselov equation and the Volterra hierarchy are
discussed.

\end{abstract}

%%%%%%%%%%%%%%%%%%%%%%%%%%%%%%%%%%%%%%%%%%%%%%%%%%%%%%%%%%%%%%%%%%%%%%%%%%
\section{Introduction.}  \label{sec-intro}
%%%%%%%%%%%%%%%%%%%%%%%%%%%%%%%%%%%%%%%%%%%%%%%%%%%%%%%%%%%%%%%%%%%%%%%%%%

In this paper I want to discuss the Backlund transformations (BTs)
for the asymmetric version of the Nizhnik-Novikov-Veselov equation
(NNVE) \cite{N,NV,DA},

\begin{equation}
\left\{
  \begin{array}{lcl}
  0 & = & - p_{t} + p_{xxx} + 6 \left( w p \right)_{x}
  \cr
  0 & = & p_{x} + w_{y}.
  \end{array}
\right.
\label{nnv-eq}
\end{equation}
The BTs which will be discussed below belong to the class of
transformations studied in \cite{BL,L,F,SY}. Contrary to the
Backlund-Darboux (or soliton-adding) transformations they do not
possess the superposition property but appear in sequences that can be
described by some discrete equations, which turn out to be
integrable as well as the original system.

I am going to start with the traditional approach: in section
\ref{sec-bt}, I present a set of relations and prove (in 
appendix A) that they indeed link different solutions of the NNVE.
Then by introducing some additional functions I rewrite these
relations as a set of bilinear equations which can be shown to
belong to the Volterra hierarchy (VH) (section \ref{sec-tovh}). To
demonstrate the relations between the NNVE and the VH, I derive the
former from the the latter (section \ref{sec-tonnv}), i.e. I show
that the NNVE can be obtained as a differential consequence of
equations of the VH. Finally, I discuss the derived BTs in the
framework of the zero curvature reprezentation (section
\ref{sec-zcr}).

%%%%%%%%%%%%%%%%%%%%%%%%%%%%%%%%%%%%%%%%%%%%%%%%%%%%%%%%%%%%%%%%%%%%%%%%%%
\section{Backlund transformations.}    \label{sec-bt}
%%%%%%%%%%%%%%%%%%%%%%%%%%%%%%%%%%%%%%%%%%%%%%%%%%%%%%%%%%%%%%%%%%%%%%%%%%

In what follows I use instead of $p$ and $w$ from (\ref{nnv-eq})
the corresponding tau-function. To introduce it we first 'solve'
the second equation of (\ref{nnv-eq}) by presenting $p$ and $w$ as
\begin{equation}
  p = - \lambda_{xy}
  \qquad
  w = \lambda_{xx}.
\end{equation}
In terms of $\lambda$ the system (\ref{nnv-eq}) becomes
\begin{equation}
  \lambda_{ty} - \lambda_{xxxy} - 6 \lambda_{xx}\lambda_{xy} = 0
\label{nnv-eq-lambda}
\end{equation}
(here I have omitted a term which does not depend on $x$ and can
be eliminated by a symmetry transform). In terms of the function
$\tau$ given by

\begin{equation}
  \tau = \exp\lambda
\end{equation}
equation (\ref{nnv-eq-lambda}) can be rewritten in the bilinear
form

\begin{equation}
  \left( D_{ty} - D_{xxxy} \right) \, \tau\cdot\tau = 0
\label{nnv-bilin}
\end{equation}
which is the simplest DKP equation, according to the
classification of Jimbo and Miwa \cite{JM}. Here the symbol $D$ 
stands for the Hirota's bilinear operators

\begin{equation}
  D_{x}^{m}D_{y}^{n} ... \; a \cdot b =
  \frac{\partial^{m}}{\partial\xi^{m}}
  \frac{\partial^{n}}{\partial\eta^{n}}
  \; ...
  \left.
  a(x+\xi, y+\eta, ... )
  b(x-\xi, y-\eta, ... )
  \right|_{\xi=\eta=...=0}.
\end{equation}

Using the tau-function $\tau$ and its logarithm $\lambda$ the
central result of the paper (the BT for the NNVE) can be
formulated as follows: if two tau-functions, $\tau$ and
$\hat\tau$, are related by
\begin{eqnarray}
  \Lambda_{t} & = &
    \Lambda_{xxx} + \Lambda_{x}^{3} + 3 \Lambda_{x}M_{xx}
    + \frac{3}{\Lambda_{x}} \lambda_{xx}\hat\lambda_{xx}
\label{bt-3}
\\
  0 & = &
    \lambda_{xy}\hat\lambda_{xy} - \Lambda_{x}
\label{bt-1}
\\
  0 & = &
    \lambda_{xxy}\hat\lambda_{xy} - \lambda_{xy}\hat\lambda_{xxy}
    + 2 \Lambda_{x}^{2} - M_{xx}
\label{bt-2}
\end{eqnarray}
where
\begin{equation}
  \Lambda = \lambda - \hat\lambda
  \qquad
  M = \lambda + \hat\lambda
\end{equation}
with
\begin{equation}
  \lambda = \ln\tau
  \qquad
  \hat\lambda = \ln\hat\tau
\end{equation}
and one of them, say $\tau$, solves (\ref{nnv-bilin}), then the
other one, $\hat\tau$, is also a solution of (\ref{nnv-bilin}).

Of course, expressions (\ref{bt-3})--(\ref{bt-2}) are rather
cumbersome and, if one wants to prove that (i) this system of
three equations for two functions is compatible or that (ii) these
relations are indeed BTs, one needs rather lengthy (though not very
difficult) calculations. In  appendix A I give a direct proof
of the statement (ii). As to the question (i) I will return to it
in section \ref{sec-tonnv} when our BT will be reformulated in a
more elegant and transparent form.

%%%%%%%%%%%%%%%%%%%%%%%%%%%%%%%%%%%%%%%%%%%%%%%%%%%%%%%%%%%%%%%%%%%%%%%%%%
\section{From NNVE to VH.}   \label{sec-tovh}
%%%%%%%%%%%%%%%%%%%%%%%%%%%%%%%%%%%%%%%%%%%%%%%%%%%%%%%%%%%%%%%%%%%%%%%%%%

The aim of this section is to show that BT
(\ref{bt-3})--(\ref{bt-2}) can be described by equations from the
VH. To do this we need to introduce some quantities which enable
us to reformulate the BTs discussed above in a more transparent
way. Consider functions
  $\sigma$, $\hat\sigma$ and $\rho$, $\hat\rho$
defined by

\begin{equation}
  \sigma =
    - \frac{ \textstyle \tau^{2} }{ \textstyle \hat\tau } \lambda_{xy}
  \qquad
  \qquad
  \hat\sigma =
    - \frac{ \textstyle \hat\tau^{2} }{ \textstyle \tau } \hat\lambda_{xy}
\label{sigma-def}
\end{equation}
and

\begin{equation}
  \rho =
    - \frac{ \textstyle \tau^{3} }{ \textstyle \hat\tau^{2} }
      \left( \lambda_{xxy} + \Lambda_{x}\lambda_{xy} \right)
  \qquad
  \qquad
  \hat\rho =
     \frac{ \textstyle \hat\tau^{3} }{ \textstyle \tau^{2} }
      \left( \hat\lambda_{xxy} - \Lambda_{x}\hat\lambda_{xy} \right).
\label{rho-def}
\end{equation}
Noting that
\begin{equation}
  \sigma\hat\sigma =
  \tau\hat\tau \;
  \lambda_{xy}\hat\lambda_{xy}  =
  \tau\hat\tau \;
  \left( \ln \frac{\tau}{\hat\tau} \right)_{x}
\label{to-1}
\end{equation}
(I have used (\ref{bt-1})) one can rewrite the last equation and
the definitions of $\rho$ and $\hat\rho$ in a similar bilinear
way:
\begin{equation}
  \begin{array}{lcl}
  D_{x} \, \sigma \cdot \tau & = & \rho \hat\tau
  \\
  D_{x} \, \tau \cdot \hat\tau & = & \sigma\hat\sigma
  \\
  D_{x} \, \hat\tau \cdot \hat\sigma & = & \tau\hat\rho.
  \end{array}
\label{pve-x}
\end{equation}
One can also obtain a set of bilinear equations involving
$y$-derivatives. Differentiating (\ref{sigma-def}) with respect to
$y$ we get
\begin{eqnarray}
  D_{y} \, \sigma \cdot \hat\tau - \tau^{2}
  & = &
  - \tau^{2} \, \Theta
\label{gtp}
  \\
  D_{y} \, \hat\sigma \cdot \tau + \hat\tau^{2}
  & = &
  - \hat\tau^{2} \, \hat\Theta
\label{gtn}
\end{eqnarray}
where
\begin{eqnarray}
  \Theta & = & \lambda_{xyy} + 2\Lambda_{y}\lambda_{xy} + 1
\label{theta-pos}
  \\
  \hat\Theta & = & \hat\lambda_{xyy} - 2\Lambda_{y}\hat\lambda_{xy} - 1
\label{theta-neg}
\end{eqnarray}
while equations (\ref{rho-def}) lead to
\begin{eqnarray}
  D_{y} \, \rho \cdot \tau - \sigma^{2}  & = &
  - \frac{\tau^{4}}{\hat\tau^{2}}
      \left( \Theta_{x} + \Lambda_{x}\Theta \right)
\label{gsp}
  \\
  D_{y} \, \hat\rho \cdot \hat\tau + \hat\sigma^{2} & = &
  \frac{\hat\tau^{4}}{\tau^{2}}
      \left( \hat\Theta_{x} - \Lambda_{x}\hat\Theta \right).
\label{gsn}
\end{eqnarray}
To calculate the quantities $\Theta$ and $\hat\Theta$ one can
obtain, by differentiating (\ref{bt-1}) and (\ref{bt-2}) with
respect to $y$, the following identities:
\begin{eqnarray}
  &&
  \hat\lambda_{xy} \Theta + \lambda_{xy} \hat\Theta = 0
\label{theta-1}
  \\&&
  \hat\lambda_{xy} \Theta_{x} - \hat\lambda_{xxy} \Theta
  - \lambda_{xy} \hat\Theta_{x} + \lambda_{xxy} \hat\Theta = 0
\label{theta-2}
\end{eqnarray}
which give
\begin{equation}
  \left( \frac{ \Theta }{ \lambda_{xy} } \right)_{x} = 0
  \qquad
  \mbox{and}
  \qquad
  \left( \frac{ \hat\Theta }{ \hat\lambda_{xy} } \right)_{x} = 0.
\label{theta-3}
\end{equation}
Hence,
\begin{equation}
  \Theta = \varphi \lambda_{xy}
  \qquad
  \mbox{and}
  \qquad
  \hat\Theta =  \hat\varphi \hat\lambda_{xy}
\end{equation}
where $\varphi$ and $\hat\varphi$ are some functions which do not
depend on $x$,
\begin{equation}
  \varphi_{x} = \hat\varphi_{x} = 0.
\label{phi-x}
\end{equation}
Returning to (\ref{theta-1}) one can conclude that
  $\hat\varphi = - \varphi$.
This leads, together with (\ref{sigma-def}), to the following
result for $\Theta$ and $\hat\Theta$:

\begin{equation}
  \Theta = - \varphi \frac{\sigma \hat\tau}{\tau^{2}}
  \qquad
  \mbox{and}
  \qquad
  \hat\Theta =  \varphi \frac{\tau \hat\sigma }{\hat\tau^{2}}.
\end{equation}
Finally, one can rewrite (\ref{gsp}), (\ref{gtp}), (\ref{gtn}) and
(\ref{gsn}) as
\begin{equation}
  \begin{array}{lcl}
  D_{y} \, \rho \cdot \tau - \varphi\, \rho \tau
  & = & \sigma^{2}
  \\
  D_{y} \, \sigma \cdot \hat\tau - \varphi\, \sigma \hat\tau
  & = & \tau^{2}
  \\
  D_{y} \, \tau \cdot \hat\sigma - \varphi\, \tau \hat\sigma
  & = & \hat\tau^{2}
  \\
  D_{y} \, \hat\tau \cdot \hat\rho - \varphi\, \hat\tau \hat\rho
  & = & \hat\sigma^{2}.
  \end{array}
\label{pve-y}
\end{equation}
One can clearly see that our six functions form a chain

\begin{equation}
  \hat\rho \to
  \hat\sigma \to
  \hat\tau \to
  \tau \to
  \sigma \to
  \rho.
\end{equation}
Indeed, after introducing the sequence of tau-functions
$\tau_{n}$,

\begin{equation}
  \begin{array}{lcl}
  \tau_{1} & = & \tau \cr
  \tau_{2} & = & \sigma \exp\left(  -\Phi \right) \cr
  \tau_{3} & = & \rho   \exp\left( -2\Phi \right) \cr
  \end{array}
  \qquad
  \begin{array}{lcl}
  \tau_{0}  & = & \hat\tau   \exp\left(  \Phi \right) \cr
  \tau_{-1} & = & \hat\sigma \exp\left( 2\Phi \right) \cr
  \tau_{-2} & = & \hat\rho   \exp\left( 3\Phi \right) \cr
  \end{array}
\end{equation}
where $\Phi$ is an antiderivative of $\varphi/2$,
\begin{equation}
  \Phi_{y} = \frac{\varphi}{2}
\end{equation}
equations (\ref{pve-x}) and (\ref{pve-y}) become
\begin{equation}
  D_{x} \, \tau_{n+1} \cdot \tau_{n} = \tau_{n+2}\tau_{n-1}
  \qquad
  n = -1, 0, 1
\label{ve-x-fin}
\end{equation}
and
\begin{equation}
  D_{y} \, \tau_{n+1} \cdot \tau_{n-1} = \tau_{n}^{2}
  \qquad
  n = -1, 0, 1,2.
\label{ve-y-fin}
\end{equation}

It can be shown that the sequence $\tau_{-2} \to ... \to \tau_{2}$
can be extended in both directions to infinity: we can introduce
functions $\tau_{n}$ for $n = \pm 3, \pm 4, ...$ in such a way
that they will satisfy (\ref{ve-x-fin}) and (\ref{ve-y-fin}),
\begin{equation}
  D_{x} \, \tau_{n+1} \cdot \tau_{n} = \tau_{n+2}\tau_{n-1}
\label{ve-x}
\end{equation}
and
\begin{equation}
  D_{y} \, \tau_{n+1} \cdot \tau_{n-1} = \tau_{n}^{2}
\label{ve-y}
\end{equation}
for every $n$ (see appendix B).

One can easily recognize in (\ref{ve-x}) the classical Volterra
chain written in the bilinear form. Indeed, the quantities $u_{n}$
defined by
\begin{equation}
  u_{n} = \frac{ \tau_{n-2}\tau_{n+1} }{ \tau_{n-1}\tau_{n} }
\label{u-def}
\end{equation}
solve the famous Volterra equation
\begin{equation}
  \partial_{x}  u_{n} =
  u_{n} \left( u_{n+1} - u_{n-1} \right)
\label{ve}
\end{equation}
(where $\partial_{x} = \partial / \partial x$). As to equations
(\ref{ve-y}), they are nothing but the simplest equations of the
negative Volterra hierarchy discussed in \cite{NVF}. Finally, the
$t$-equation (\ref{bt-3}) of our BT is, in the terms of the
tau-functions $\tau_{n}$, the \emph{third} equation of the
classical (positive) VH,
\begin{equation}
  \left( D_{t} - D_{xxx} \right) \tau_{n} \cdot \tau_{n-1} =
  3 \tau_{n-3}\tau_{n+2}.
\label{ve-t}
\end{equation}

Now let us compare our BTs with the standard ones which can be
derived using the Hirota's bilinear approach as follows. Denoting
the left-hand side of (\ref{nnv-bilin}) by $E$,
\begin{equation}
  E(\tau) = \left( D_{ty} - D_{xxxy} \right) \, \tau \cdot \tau
\label{def-E}
\end{equation}
one can decompose the difference
  $E(\tau) \, \widetilde\tau^{2} - \tau^{2} E(\widetilde\tau)$
as
\begin{equation}
  E(\tau) \, \widetilde\tau^{2} - \tau^{2} E(\widetilde\tau) =
  2 D_{y} \;
    \left[
      \left(D_{t} - D_{xxx} \right) \tau \cdot \widetilde\tau
    \right]
    \cdot
    \tau \widetilde\tau
  +
  6 D_{x} \;
    \left( D_{xy} \, \tau \cdot \widetilde\tau \right)
    \cdot
    \left( D_{x} \, \tau \cdot \widetilde\tau \right).
\end{equation}
This implies that, if functions $\tau$ and $\widetilde\tau$ are
related by
\begin{equation}
  \left\{
  \begin{array}{l}
    \left(D_{t} - D_{xxx} - 3\alpha \right) \tau \cdot \widetilde\tau
    = 0
    \qquad \alpha_{y}=0
    \cr
    \left(D_{xy} - \beta D_{x} \right) \tau \cdot \widetilde\tau
    = 0
    \qquad \qquad \beta_{x}=0
  \end{array}
  \right.
\label{bls-1}
\end{equation}
and $E(\tau) = 0$, then $E(\widetilde\tau) = 0$ as well, i.e. that
system (\ref{bls-1}) describes a BT.

It was noted in the introduction that the BTs of the type
discussed in this paper differ essentially from the usual BTs
(\ref{bls-1}). This fact manifests itself in many aspects. Usually
applications of BTs (\ref{bls-1}) lead from simple solutions to
more complicated ones. The most bright example is the following.
Let us apply (\ref{bls-1}) to the trivial tau-function $\tau=1$.
By solving the corresponding linear equations we come to the
tau-function $\widetilde\tau$ which is a generalization of the
one-soliton tau-function that can be obtained, say, by the
Hirota's method. At the same time the simplest (vacuum) solution
of the Volterra equations remains trivial (in the sense that it
gives the zero solution of the NVVE) for all the values of $n$.
The same can be said about more complex situations. BTs
(\ref{bls-1}) at every step change the structure of solutions (add
a soliton), while a shift of the Volterra index, $n \to n \pm 1$,
usually leads to less essential changes. The only known exception
is the determiant (Hankel, Wroskian etc) solutions where the index
$n$ is related to the size of the matrix. However these solutions
have not been extended yet to satisfy the negative equations of the
VH, so I cannot discuss them in the context of the NVVE.

The relationships between the BTs of both the kinds have been
studied for the (1+1)-dimensional integrable systems such as the
modified KdV or nonlinear Schrodinger equations (see, e.g, book by
Newell \cite{Newell}). In the most transparent way the difference
between them can be exposed in the framework of the inverse
scattering approach. The soliton-adding BTs similar to
(\ref{bls-1}) are related to the Darboux transformations of the
associated linear problem: they add zeroes to the corresponding
scattering data. That is why they are often called
Backlund-Darboux transformations. Transformations similar to those
proposed in this paper, which are described by integrable chains
(for example, in the case of the nonlinear Schrodinger equation it
is the Toda chain), are of different kind. They are related to the
Schlesinger transformations of the scattering problem and change
the asymptotics (monodromy) of the fundamental solutions as
functions of the spectral parameter (see, e.g., chapter 5 of
\cite{Newell}). However the inverse scattering approach for the NVVE is
out of the scope of this paper and more elaborated discussion of
this question needs further studies.

However, in some cases the relationships between both kinds of 
transformations are rather transparent.
One can show using equations (\ref{ve-x}) and (\ref{ve-y})
that Volterra tau-functions solve (in the case of the proper
boundary conditions)
\begin{equation}
  D_{xy} \, \tau_{n} \cdot \tau_{n-1} +
  D_{y}  \, \tau_{n+1} \cdot \tau_{n-2} = 0.
\end{equation}
Taking this equation together with the third Volterra equation
(\ref{ve-t}) one can show that Volterra tau-functions satisfy the
system
\begin{equation}
  \left\{
  \begin{array}{l}
    \left(D_{t} - D_{xxx} \right) \tau_{n} \cdot \tau_{n-1}
    - 3 \tau_{n-3}\tau_{n+2}
    = 0
    \cr
    D_{xy} \, \tau_{n}   \cdot \tau_{n-1} +
    D_{y} \,  \tau_{n+1} \cdot \tau_{n-2}
    = 0
  \end{array}
  \right.
\label{bls-2}
\end{equation}
which can be reduced to the one similar to (\ref{bls-1}). Indeed,
in the three-periodic case
\begin{equation}
  \tau_{n+3} = \tau_{n}
\label{bls-period}
\end{equation}
equations (\ref{bls-2}) become
\begin{equation}
  \left\{
  \begin{array}{l}
    \left(D_{t} - D_{xxx} - 3 \right) \tau_{n} \cdot \tau_{n-1}
    = 0
    \cr
      D_{xy} \; \tau_{n} \cdot \tau_{n-1}
    = 0.
  \end{array}
  \right.
\label{bls-3}
\end{equation}
(note that (i) the VH $\to$ NVVE correspondence is valid for
\emph{any} solutions of the VH and that (ii) periodic reductions
are compatible with the Volterra equations). Comparing
(\ref{bls-3}) with (\ref{bls-1}) one can easily note that
(\ref{bls-3}) coincide with (\ref{bls-1}) with $\alpha=1$ and
$\beta = 0$. So, one can conclude that the BTs described by
the three-periodic Volterra equations are a particular case of the 
classical BTs (\ref{bls-1}). This situation, when Schlesinger 
transformations turn out to be some special (sometimes singular) 
cases of the Backlund-Darboux transformations, has been already 
discussed in the literature (see, e.g., \cite{Leble,DMAMP}).
Of course the calculations presented above cannot give an exhaustive 
analysis of the relations between these BTs. I repeat, they are restricted 
to the three-periodic case, when the Volterra chain, 
I would like to recall, is nothing but 
a Painleve, namely the PIV, equation which is known to posses a 
number of peculiar features. Thus, the question of how to construct 
the BTs described by the VH of the Backlund-Darboux transformations 
remains to be settled and surely deserves further studies 
(probably in a more general framework).

%%%%%%%%%%%%%%%%%%%%%%%%%%%%%%%%%%%%%%%%%%%%%%%%%%%%%%%%%%%%%%%%%%%%%%%%%%
\section{From VH to NNVE.}   \label{sec-tonnv}
%%%%%%%%%%%%%%%%%%%%%%%%%%%%%%%%%%%%%%%%%%%%%%%%%%%%%%%%%%%%%%%%%%%%%%%%%%

The way from the VH to the NNVE is more straightforward than the
calculations described in the previous sections: one has to write
down three equations of the extended VH and show that any of their
common solution also satisfies equations (\ref{nnv-eq}) or
(\ref{nnv-bilin}).

The VH is an infinite set of differential-difference equations
compatible with the discrete problem
\begin{equation}
  \psi_{n-1} - \psi_{n} + \zeta u_{n} \psi_{n+1} = 0,
  \qquad
  u_{n} = \frac{\tau_{n-2}\tau_{n+1}}{\tau_{n-1}\tau_{n}}.
\label{zcr-1}
\end{equation}
The classical (or 'positive') Volterra equations are evolution
equations $\partial u_{n} / \partial t_{j} = F_{n}^{(j)}$, two of
which have already been written down (see (\ref{ve}) and
(\ref{ve-t})). Another type of the Volterra eqautions, which are
non-local and which when taken together form the 'negative' VH, were
discussed in \cite{NVF}. The simplest of them is (\ref{ve-y}). The
very important point is the fact that all equations of the
extended VH (both the classical and the 'negative' ones) \emph{are
compatible}. So, we can consider them simultaneously as one
infinite system. We can think of $u_{n}$ (or $\tau_{n}$) as
functions of an infinite number of times,
  $\tau_{n} =
  \tau_{n}\left( t_{1}, t_{2}, ..., \bar t_{1}, ... \right)$,
where the dependence on $t_{j}$ ($\bar t_{j}$) is determined by
the $j$th 'positive' ('negative') Volterra equation.

Hereafter I will restrict myself to the finite subsystem of the
VH, consisting of the equations mentioned above: the first and the
third equations of the 'positive' subhierarchy (the corresponding
times will be denoted by $x$ and $t$) and the first 'negative'
equation (with $y$ used instead of $\bar t_{1}$). So, we will deal
with the system
\begin{eqnarray}
  D_{x} \, \tau_{n} \cdot \tau_{n-1} & = & \tau_{n-2}\tau_{n+1}
\label{syst-x}
\\
  \left( D_{t} - D_{xxx} \right) \tau_{n} \cdot \tau_{n-1} & = &
  3 \tau_{n-3}\tau_{n+2}
\label{syst-t}
\\
  D_{y} \, \tau_{n+1} \cdot \tau_{n-1} & = & \tau_{n}^{2}
\label{syst-y}
\end{eqnarray}
and before proceeding further 
I would like to return to the question
of the compatibility of system (\ref{bt-3})--(\ref{bt-2}). It
turns out that we do not need to prove this fact separately:
system (\ref{bt-3})--(\ref{bt-2}) is equivalent to
(\ref{syst-x})--(\ref{syst-y}) which is a part of the VH, while
consistency of the later has already been established
(compatibility of equations of an integrable hierarchy, or
commutativity of the corresponding flows, is one of the
ingredients of its integrability).

The main result of this section, the transition from the VH to
NNVE, can be achieved by the following simple calculations.
Returning from the Hirota's bilinear differential operators to
usual ones, one can present equation (\ref{syst-t}) with the help
of (\ref{syst-x}) as
\begin{equation}
  \left( \partial_{t} - \partial_{xxx} \right)
    \ln \frac{\tau_{n}}{\tau_{n-1}} =
   3 u_{n+1}u_{n}u_{n-1} +
   3 u_{n+1}u_{n}^{2} +
   3 u_{n}^{2}u_{n-1} +
   u_{n}^{3}
\end{equation}
which gives for the quantity $p_{n}$,
\begin{equation}
  p_{n} = \frac{\tau_{n-1}\tau_{n+1}}{\tau_{n}^{2}}
\label{p-def}
\end{equation}
the identity
\begin{equation}
  \left( \partial_{t} - \partial_{xxx} \right) p_{n} =
  6 \, \partial_{x} \left( w_{n}p_{n} \right)
\label{vh-p}
\end{equation}
where
\begin{equation}
  w_{n} = \frac{\tau_{n-2}\tau_{n+2}}{\tau_{n}^{2}} =
  u_{n}u_{n+1}.
\label{w-def}
\end{equation}
On the other hand, (\ref{syst-y}) leads to
\begin{equation}
  \partial_{y} w_{n} = p_{n} \left( u_{n} - u_{n+1} \right)
\end{equation}
Applying (\ref{syst-x}) again one can easily show that the
right-hand side of the last equation is nothing but
$-\partial_{x}p_{n}$. So,
\begin{equation}
  \partial_{y} w_{n} + \partial_{x} p_{n} = 0.
\label{vh-w}
\end{equation}
Comparing (\ref{vh-p}) and (\ref{vh-w}) with (\ref{nnv-eq}) one
can see that \textit{for any} $n$ functions $p_{n}$ and $w_{n}$
solve the NNVE.

Using the fact that
  $w_{n} = \partial_{xx}\ln\tau_{n}$
and
  $p_{n}=-\partial_{xy}\ln\tau_{n}$
(these formulae can be derived from (\ref{syst-x}) and
(\ref{syst-y}) after neglecting some unessential constants, which
can be incorporated in the definition of $\tau_{n}$) we can
reformulate this result as follows: for each $n$ the tau-function
of the VH, $\tau_{n}$, is a solution of the bilinear NNVE
(\ref{nnv-bilin}). In the context of the NNVE the meaning of the
Volterra index $n$ is clear: $n$ is the number of our solution in
the sequence of the BTs discussed in the previous section.

%%%%%%%%%%%%%%%%%%%%%%%%%%%%%%%%%%%%%%%%%%%%%%%%%%%%%%%%%%%%%%%%%%%%%%%%%%
\section{BTs and zero-curvature representation.}   \label{sec-zcr}
%%%%%%%%%%%%%%%%%%%%%%%%%%%%%%%%%%%%%%%%%%%%%%%%%%%%%%%%%%%%%%%%%%%%%%%%%%

So far we have discussed the BTs in terms of solutions of our
nonlinear equation only: both (\ref{bt-3})--(\ref{bt-2}) and
(\ref{bls-1}) are expressions which establish some links between
different solutions of the NNVE. However, the structure of BTs for
nonlinear integrable systems becomes more transparent when
expressed in terms of the solutions of auxiliary linear problems. For
example the NNVE can be presented as the compatibility condition
for the system
\begin{equation}
\left\{
  \begin{array}{lcl}
  \varphi_{xy} & = & 2 p \varphi
  \cr
  \varphi_{t} & = &  \varphi_{xxx} + 6 w \varphi_{x}
  \end{array}
\right. \label{zcr-trad}
\end{equation}
(the so-called zero-curvature representation) and transfom
(\ref{bls-1}) $\tau \to \widetilde\tau$ with $\beta = 0$,
\begin{equation}
  D_{xy} \, \tau \cdot \widetilde\tau = 0
\end{equation}
leads to the following transformation of $\varphi$
\begin{equation}
  \varphi \to \widetilde\varphi: \qquad
  \left\{
  \begin{array}{lcl}
    \widetilde\varphi_{x} - \Lambda_{x}\widetilde\varphi =
    - \varphi_{x} - \Lambda_{x}\varphi
  \cr
    \widetilde\varphi_{y} - \Lambda_{y}\widetilde\varphi =
    \phantom{-} \varphi_{y} + \Lambda_{y}\varphi
  \end{array}
  \right.
  \qquad
  \Lambda = \ln\frac{\tau}{\widetilde\tau}
\end{equation}
(these transformations are known as Loewner transformations
\cite{Lo} and were discussed e.g. in \cite{KR,SR}).  I cannot
present simple formulae for the Volterra sequence of the BTs in
the framework of auxiliary problem (\ref{zcr-trad}). It turns out
that to describe the BTs of this paper it is more convenient to
use another linear problem associated with the NNVE which can be
derived from the zero-curvature representation of the VH.

It can be shown that evolution of the functions $\psi_{n}$ 
from (\ref{zcr-1}) with respect to the flows described 
by (\ref{syst-x})--(\ref{syst-y}) can be written as
\begin{eqnarray}
  \partial_{x} \psi_{n} & = &
    u_{n} \left( \psi_{n+1} - \psi_{n} \right)
\label{dpsi-x}
\\[2mm]
  \partial_{y} \psi_{n} & = &
    \frac{1}{p_{n-1}} \left( \psi_{n-1} - \psi_{n} \right)
\label{dpsi-y}
\\
  \partial_{t} \psi_{n} & = &
    u_{n}u_{n+1}u_{n+2} \psi_{n+3} +
    \alpha_{n} \psi_{n+2} +
    \beta_{n} \psi_{n+1} +
    \gamma_{n} \psi_{n}
\label{dpsi-t}
\end{eqnarray}
where $u_{n}$ and $p_{n}$ are defined in (\ref{u-def}) and
(\ref{p-def}), while $\alpha_{n}$, $\beta_{n}$ and $\gamma_{n}$
are some coefficients not written here explicitly. Equations
(\ref{dpsi-x}) and (\ref{dpsi-y}) lead to
\begin{equation}
  \partial_{xy} \psi_{n} =
  p_{n} \left( - \psi_{n-1} + 2\psi_{n} - \psi_{n+1} \right)
\end{equation}
which can be transformed to
\begin{equation}
  \partial_{xy} \psi_{n} +
  \frac{1}{p_{n-1}} \partial_{x} \psi_{n} +
  u_{n} \partial_{y} \psi_{n} = 0.
\label{aux-xy}
\end{equation}
At the same time equation (\ref{dpsi-t}) can be presented as
\begin{equation}
  \partial_{t} \psi_{n} =
  \partial_{xxx} \psi_{n} +
  3 \left(u_{n-1}+u_{n}\right) \, \partial_{xx} \psi_{n} +
  3\left[ \left(u_{n-1}+u_{n}\right)^{2} + w_{n} \right] \, \partial_{x} \psi_{n}
\label{aux-t}
\end{equation}
where $w_{n}=u_{n}u_{n+1}$ (see (\ref{w-def})). Note that the last
two equations can be rewritten in the 'one-site' form
\begin{eqnarray}
  \partial_{xy} \psi_{n} & = &
    \mu_{n} \partial_{x} \psi_{n} +
    \nu_{n} \partial_{y} \psi_{n}
\label{dpsi-xy}
\\
  \partial_{t} \psi_{n} & = &
  \partial_{xxx} \psi_{n} +
  3a_{n} \, \partial_{xx} \psi_{n} +
  3\left( a_{n}^{2} + w_{n} \right) \, \partial_{x} \psi_{n}
\label{dpsi-evol}
\end{eqnarray}
where
\begin{equation}
  \mu_{n} = \frac{ \partial_{y} h_{n} }{ h_{n} },
  \qquad
  \nu_{n} = \frac{ \partial_{xy} h_{n} }{ 2 \, \partial_{y} h_{n} },
  \qquad
  a_{n} = - \frac{ \partial_{x} h_{n} }{ h_{n} }
\end{equation}
with
\begin{equation}
  h_{n} = \frac{ \tau_{n-2} }{ \tau_{n} }.
\end{equation}

Equations (\ref{dpsi-xy}) and (\ref{dpsi-evol}) can be used as a
zero-curvature representation of the NNVE, different from the
traditional one given by (\ref{zcr-trad}). Indeed, one can check
by straightforward calculatioins that the compatibility condition
for the system
\begin{eqnarray}
  \psi_{xy} & = & \mu \psi_{x} + \nu \psi_{y}
\label{lp-xy}
\\
  \psi_{t} & = &
    \psi_{xxx} +
    3 a \psi_{xx} +
    3\left( a^{2} + w \right) \psi_{x}
\label{lp-t}
\end{eqnarray}
where
\begin{equation}
  \mu = \frac{ h_{y} }{ h },
  \qquad
  \nu = \frac{ h_{xy} }{ 2 h_{y} },
  \qquad
  a = - \frac{ h_{x} }{ h }
\end{equation}
can be reduced to the system (\ref{nnv-eq}) for the quantities $w$
and $p=\mu\nu$.

In terms of this zero-curvature representation it is clearly seen
that the Volterra sequence of BTs discussed in this paper can be
constructed by means of rising/lowering operators
$\psi_{n}\to\psi_{n \pm 1}$ given by (\ref{dpsi-x}) and
(\ref{dpsi-y}):
\begin{eqnarray}
  \psi_{n+1} & = &
    \left( \frac{1}{u_{n}} \partial_{x} + 1 \right) \psi_{n}
\\
  \psi_{n-1} & = &
    \left( p_{n-1} \partial_{y} + 1 \right) \psi_{n}.
\end{eqnarray}
Thus our sequence of BTs for NNVE is generated by the iteration of
the Laplace-Darboux transformations for linear problem
(\ref{lp-xy}) and (\ref{lp-t}).

%%%%%%%%%%%%%%%%%%%%%%%%%%%%%%%%%%%%%%%%%%%%%%%%%%%%%%%%%%%%%%%%%%%%%%%%%%
\section{Conclusion.}
%%%%%%%%%%%%%%%%%%%%%%%%%%%%%%%%%%%%%%%%%%%%%%%%%%%%%%%%%%%%%%%%%%%%%%%%%%

In this paper I have derived the BTs for the NNVE and have shown
that these BTs can be iterated and that the resulting sequence can
be described by the Volterra equations. The same result can be
reformulated in a more general way: the NNVE can be embedded in
the extended Volterra hierarchy by presenting it as a result of
the combined action of the Volterra flows.

At the end I want to give the following remark. Among the three
Volterra flows we used there were two 'positive' ones,
$\partial/\partial t_{1}$ and $\partial/\partial t_{3}$. So, an
interesting question is about the role of the 'skipped' second
Volterra flow, $\partial/\partial t_{2}$. In terms of the NNVE, 
$\partial/\partial t_{2}$ can be viewed as some nonlocal symmetry.
This symmetry could be used to derive the BTs, but I preferred
more standard approach of introducing additional tau-functions
($\sigma$, $\rho$, ...) instead of introducing additional
independent variables ($t_{2}$ and other). However, it should be
noted that the second strategy is already known and had been
shown, say, in \cite{LLSW} to be rather useful in a wide range of
situations. The question of $t_{2}$-dependence is also interesting
because it leads us to the Kadomtsev-Petviashvili (KP) equation.
Indeed, it can be shown that any tau-function $\tau_{n}$ of the
Volterra hierarchy solves also

\begin{equation}
  \left(
    4D_{1}D_{3} - 3 D_{2}^{2} - D_{1}^{4}
  \right) \, \tau_{n}\cdot\tau_{n} = 0
\end{equation}
where $D_{j}$ are the Hirota operators corresponding to $t_{j}$,
which means that $\tau_{n}$ is a tau function of the KP equation
as well (or, in other words, that the KP equation can be embedded
in the VH). So, the NNVE can be considered as a (nonlocal)
symmetry of the KP equation. This fact can enlarge the area of
application of the former and clarify its place among other
integrable partial differential equations.

%%%%%%%%%%%%%%%%%%%%%%%%%%%%%%%%%%%%%%%%%%%%%%%%%%%%%%%%%%%%%%%%%%%%%%%%%%
\section*{Acknowledgements.}
%%%%%%%%%%%%%%%%%%%%%%%%%%%%%%%%%%%%%%%%%%%%%%%%%%%%%%%%%%%%%%%%%%%%%%%%%%

I wish to thank A.V. Mikhailov for helpful advice and comments.
This work is supported by Ministerio de Educaci\'on, Cultura y
Deporte of Spain under grant SAB2000-0256.

%%%%%%%%%%%%%%%%%%%%%%%%%%%%%%%%%%%%%%%%%%%%%%%%%%%%%%%%%%%%%%%%%%%%%%%%%%
\section*{Appendix A.}
%%%%%%%%%%%%%%%%%%%%%%%%%%%%%%%%%%%%%%%%%%%%%%%%%%%%%%%%%%%%%%%%%%%%%%%%%%
\renewcommand{\theequation}{A.\arabic{equation}}

A proof of the fact that (\ref{bt-3})--(\ref{bt-2}) are indeed the
BTs can be given as follows. For the quantity $E$ defined by
(\ref{def-E}),

\begin{equation}
  E(\tau) = \left( D_{ty} - D_{xxxy} \right) \, \tau \cdot \tau,
\end{equation}
one can get, using explicit expressions for the Hirota's bilinear
operators, that

\begin{equation}
  \frac{ E(\tau) }{6\tau^2} - \frac{ E(\hat\tau) }{6\hat\tau^2} =
  \frac{1}{3} \Lambda_{ty} - \frac{1}{3} \Lambda_{xxxy}
  - \Lambda_{xy}M_{xx} - \Lambda_{xx}M_{xy}.
\end{equation}
Substitution of $\Lambda_{t}$ from (\ref{bt-3}) gives

\begin{equation}
  \frac{ E(\tau) }{6\tau^2} - \frac{ E(\hat\tau) }{6\hat\tau^2} =
  \left( \frac{\lambda_{xx}\hat\lambda_{xx}}{\Lambda_{x}} \right)_{y}
  + \Lambda_{x}M_{xxy} - \Lambda_{xx}M_{xy} + \Lambda_{x}^{2} \Lambda_{xy}.
\label{proof-main}
\end{equation}
To proceed further we derive from (\ref{bt-1}) and (\ref{bt-2})
expressions for $\lambda_{xxy}$ and $\hat\lambda_{xxy}$. After
differentiating with respect to $x$ equation (\ref{bt-1}) becomes

\begin{equation}
  \lambda_{xxy} \hat\lambda_{xy} +
  \lambda_{xy}  \hat\lambda_{xxy} -
  \Lambda_{xx}
  = 0.
\end{equation}
Adding/subtracting (\ref{bt-2}) to/from this identity one can get
\begin{eqnarray}
  0 & = &
  \lambda_{xxy} \hat\lambda_{xy} + \Lambda_{x}^{2} - \lambda_{xx}
\label{proof-1}
\\
  0 & = &
  \lambda_{xy} \hat\lambda_{xxy} - \Lambda_{x}^{2} + \hat\lambda_{xx}
\label{proof-2}
\end{eqnarray}
which leads, after multiplying (\ref{proof-1}) by $\lambda_{xy}$,
(\ref{proof-2}) by $\hat\lambda_{xy}$ and using (\ref{bt-1}), to
\begin{eqnarray}
  \Lambda_{x} \lambda_{xxy} & = &
  \lambda_{xy} \left( \lambda_{xx} - \Lambda_{x}^{2} \right)
\label{proof-3}
\\
  \Lambda_{x} \hat\lambda_{xxy} & = &
  \hat\lambda_{xy} \left( \Lambda_{x}^{2} - \hat\lambda_{xx} \right).
\label{proof-4}
\end{eqnarray}
Now one can calculate
  $\left( \lambda_{xx}\hat\lambda_{xx} / \Lambda_{x} \right)_{y}$:

\begin{eqnarray}
  \left( \frac{\lambda_{xx}\hat\lambda_{xx}}{\Lambda_{x}} \right)_{y}
  & = &
  \frac{\lambda_{xx}}{\Lambda_{x}} \hat\lambda_{xxy} +
  \frac{\hat\lambda_{xx}}{\Lambda_{x}} \lambda_{xxy} -
  \frac{\lambda_{xx}\hat\lambda_{xx}}{\Lambda_{x}^{2}} \Lambda_{xy}
\\
  & = &
    \frac{ \lambda_{xx}\hat\lambda_{xy}}{\Lambda_{x}^{2}}
    \left( \Lambda_{x}^{2} - \hat\lambda_{xx} \right)
  + \frac{ \lambda_{xy}\hat\lambda_{xx}}{\Lambda_{x}^{2}}
    \left( \lambda_{xx} - \Lambda_{x}^{2} \right)
  - \frac{ \Lambda_{xy} }{ \Lambda_{x}^{2} }
    \lambda_{xx}\hat\lambda_{xx}
\\
  & = &
  \lambda_{xx}\hat\lambda_{xy} - \lambda_{xy}\hat\lambda_{xx}.
\label{proof-5}
\end{eqnarray}

On the other hand, summarizing (\ref{proof-3}) and (\ref{proof-4})
one can get

\begin{eqnarray}
  \Lambda_{x}M_{xxy} & = &
  \lambda_{xx}\lambda_{xy} -
  \hat\lambda_{xx}\hat\lambda_{xy} -
  \Lambda_{x}^{2}\Lambda_{xy}
\\ & = &
  \lambda_{xx} \left( M_{xy} - \hat\lambda_{xy} \right) +
  \hat\lambda_{xx} \left( \lambda_{xy} - M_{xy} \right) -
  \Lambda_{x}^{2} \Lambda_{xy}
\end{eqnarray}
which leads to

\begin{equation}
  \Lambda_{x}M_{xxy} - \Lambda_{xx}M_{xy} +  \Lambda_{x}^{2}\Lambda_{xy}
  =
  \lambda_{xy} \hat\lambda_{xx} - \lambda_{xx} \hat\lambda_{xy}.
\label{proof-6}
\end{equation}
Comparing (\ref{proof-5}) and (\ref{proof-6}) one can conclude
that the right-hand side of (\ref{proof-main}) equals to zero.
This means that

\begin{equation}
  E(\tau)=0
  \qquad
  \Rightarrow
  \qquad
  E(\hat\tau)=0
\end{equation}
if $\tau$ and $\hat\tau$ satisfy (\ref{bt-3})--(\ref{bt-2}). This
completes the proof of the fact that relations
(\ref{bt-3})--(\ref{bt-2}) are indeed a BT for the NNVE.

%%%%%%%%%%%%%%%%%%%%%%%%%%%%%%%%%%%%%%%%%%%%%%%%%%%%%%%%%%%%%%%%%%%%%%%%%%
\section*{Appendix B}
%%%%%%%%%%%%%%%%%%%%%%%%%%%%%%%%%%%%%%%%%%%%%%%%%%%%%%%%%%%%%%%%%%%%%%%%%%
\renewcommand{\theequation}{B.\arabic{equation}}
\setcounter{equation}{0}

Here I discuss a proof of the fact that finite system
(\ref{ve-x-fin}), (\ref{ve-y-fin}) can be extended to infinity.
Consider first the bilinear quantities $V_{n}$, $\overline{V}_{n}$
and $C_{n}$ defined by
\begin{eqnarray}
  V_{n} & = &
    D_{x} \, \tau_{n} \cdot \tau_{n-1} - \tau_{n+1}\tau_{n-2}
  \\
  \overline{V}_{n} & = &
    D_{y} \, \tau_{n+1} \cdot \tau_{n-1} - \tau_{n}^{2}
  \\
  C_{n} & = &
    \frac{1}{2} D_{xy} \, \tau_{n} \cdot \tau_{n} + \tau_{n+1}\tau_{n-1}
\end{eqnarray}
From the definition of the Hirota's operators one can derive by
simple algebra the fourth-order identities
\begin{equation}
    \tau_{n}\tau_{n-1}V_{n+1}
  - \tau_{n+1}\tau_{n}V_{n}
  - D_{x} \, \overline{V}_{n} \cdot \tau_{n+1}\tau_{n-1}
  + \tau_{n-1}^{2} C_{n+1}
  - \tau_{n+1}^{2} C_{n-1}
  = 0
\label{fl-p}
\end{equation}
and
\begin{equation}
    \tau_{n+1}\tau_{n-1}\overline{V}_{n+1}
  - \tau_{n+2}\tau_{n}\overline{V}_{n}
  + D_{y} \, V_{n+1} \cdot \tau_{n+1}\tau_{n}
  - \tau_{n}^{2} C_{n+1}
  + \tau_{n+1}^{2} C_{n}
  = 0
\label{fl-n}
\end{equation}
that are satisfied by \emph{any} sequences of $\tau_{n}$'s. My aim
now is to show that if $V_{n}=\overline{V}_{n}=0$ for a
sufficiently large number of sequential values of $n$, then all of
the $V_{n}$ and $\overline{V}_{n}$ are equal to zero as well. To
do that I rewrite (\ref{fl-p}) and (\ref{fl-n}) as
\begin{eqnarray}
  &&
  \frac{1}{p_{n}} \left( Y_{n+1} - Y_{n} \right)
  - \partial_{x} \overline{Y}_{n}
  + Z_{n+1} - Z_{n-1}
  = 0
\\
  &&
  u_{n} \left( \overline{Y}_{n+1} - \overline{Y}_{n} \right)
  + \partial_{x} Y_{n+1}
  - Z_{n+1} + Z_{n}
  = 0
\end{eqnarray}
where
\begin{equation}
  Y_{n} = \frac{V_{n}}{\tau_{n-1}\tau_{n}},
  \qquad
  \overline{Y}_{n} = \frac{\overline{V}_{n}}{\tau_{n-1}\tau_{n+1}},
  \qquad
  Z_{n} = \frac{C_{n}}{\tau_{n}^{2}}
\end{equation}
(I presume that none of $\tau_{n}$ is equal to zero) from which it
follows that
\begin{equation}
  \frac{1}{p_{n}} \left( Y_{n+1} - Y_{n} \right)
  + \partial_{x} \left( Y_{n+1} + Y_{n} \right)
  =
  u_{n+1} \left( \overline{Y}_{n} - \overline{Y}_{n+1} \right)
  + u_{n} \left( \overline{Y}_{n-1} - \overline{Y}_{n} \right)
  + \partial_{x} \overline{Y}_{n}
\end{equation}
or
\begin{equation}
  \overline{Y}_{n+1} =
  lin \left( Y_{n+1}, Y_{n}, \overline{Y}_{n}, \overline{Y}_{n-1} \right)
\label{iter}
\end{equation}
where $lin(...)$ is a linear combination of its arguments (and
their derivatives).

Now we can return to the Volterra equations. The finite Volterra
system (\ref{ve-x-fin}) and (\ref{ve-y-fin}) can be written as
\begin{equation}
  Y_{0} = Y_{1} = Y_{2} = 0
  \qquad  \mbox{and} \qquad
  \overline{Y}_{-1} = \overline{Y}_{0} = \overline{Y}_{1} = \overline{Y}_{2} = 0
\end{equation}
So, if we \emph{define} $\tau_{4}$ as
  $\tau_{4} = \left( D_{x} \, \tau_{3} \cdot \tau_{2} \right) / \tau_{1}$
(which means that $Y_{3}=0$), then by virtue of (\ref{iter}) we
get $\overline{Y}_{3}=0$. Repeating this procedure we can define
an infinite set of $\tau_{n}$ in such a way that
\begin{equation}
  Y_{n} = \overline{Y}_{n} = 0
  \qquad \mbox{for} \qquad
  n \geq 0
\end{equation}

It is also possible to \emph{define} tau-functions for $n \leq -3$
to ensure vanishing of all $Y_{n}$ and $\overline{Y}_{n}$ for
$n<0$. This means that these tau-functions will be solutions of
the first positive and first negative Volterra equations:
\begin{equation}
  V_{n} = 0,
  \quad
  \overline{V}_{n} = 0
  \qquad \mbox{for} \qquad
  n = 0, \pm 1, \pm 2, ...
\end{equation}

In a similar way one can consider the chains of identities for the
bilinear combinations of tau-functions which generate the
$t$-equation (\ref{ve-t}) of the VH.

%%%%%%%%%%%%%%%%%%%%%%%%%%%%%%%%%%%%%%%%%%%%%%%%%%%%%%%%%%%%%%%%%%%%%%%%%%

%%%%%%%%%%%%%%
\end{document}